\documentclass[10pt,twocolumn,twoside]{IEEEtran}

\renewcommand{\theenumii}{.\arabic{enumii}}

\usepackage{cite}
\usepackage{graphicx}  
\usepackage{amsmath}   
\usepackage{subfigure}
\usepackage{algorithm}
\usepackage{algorithmic}
\usepackage{amssymb}
\usepackage{flushend}
\usepackage{color}

\newcommand{\be}{\begin{equation}}
\newcommand{\ee}{\end{equation}}
\newcommand{\ba}{\begin{array}}
\newcommand{\ea}{\end{array}}
\newcommand{\bea}{\begin{eqnarray}}
\newcommand{\eea}{\end{eqnarray}}

\newcommand{\herm}{^{\mbox{\scriptsize H}}}

\newcommand{\vbar}{\raisebox{.17ex}{\rule{.04em}{1.35ex}}}
\newcommand{\vbarind}{\raisebox{.01ex}{\rule{.04em}{1.1ex}}}
\newcommand{\R}{\ifmmode {\rm I}\hspace{-.2em}{\rm R} \else ${\rm I}\hspace{-.2em}{\rm R}$ \fi}
\newcommand{\T}{\ifmmode {\rm I}\hspace{-.2em}{\rm T} \else ${\rm I}\hspace{-.2em}{\rm T}$ \fi}
\newcommand{\N}{\ifmmode {\rm I}\hspace{-.2em}{\rm N} \else \mbox{${\rm I}\hspace{-.2em}{\rm N}$} \fi}
\newcommand{\B}{\ifmmode {\rm I}\hspace{-.2em}{\rm B} \else \mbox{${\rm I}\hspace{-.2em}{\rm B}$} \fi}
\newcommand{\Hil}{\ifmmode {\rm I}\hspace{-.2em}{\rm H} \else \mbox{${\rm I}\hspace{-.2em}{\rm H}$} \fi}
\newcommand{\C}{\ifmmode \hspace{.2em}\vbar\hspace{-.31em}{\rm C} \else \mbox{$\hspace{.2em}\vbar\hspace{-.31em}{\rm C}$} \fi}
\newcommand{\Cind}{\ifmmode \hspace{.2em}\vbarind\hspace{-.25em}{\rm C} \else \mbox{$\hspace{.2em}\vbarind\hspace{-.25em}{\rm C}$} \fi}
\newcommand{\Q}{\ifmmode \hspace{.2em}\vbar\hspace{-.31em}{\rm Q} \else \mbox{$\hspace{.2em}\vbar\hspace{-.31em}{\rm Q}$} \fi}
\newcommand{\Z}{\ifmmode {\rm Z}\hspace{-.28em}{\rm Z} \else ${\rm Z}\hspace{-.28em}{\rm Z}$ \fi}

\renewcommand{\vec}[1]{\bf{#1}}     

\IEEEoverridecommandlockouts

\begin{document}

\title{\textcolor{black}{Distributed Coordinated Beamforming for Multi-cell Multigroup Multicast Systems}}

\author{Harri Pennanen, Dimitrios Christopoulos, Symeon Chatzinotas and  Bj\"orn Ottersten \\
SnT - securityandtrust.lu, University of Luxembourg \\
e-mail: \{harri.pennanen, dimitrios.christopoulos, symeon.chatzinotas, bjorn.ottersten\}@uni.lu}

\maketitle

\begin{abstract}
This paper considers coordinated multicast beamforming in a multi-cell wireless network. Each multiantenna base station (BS) serves multiple groups of single antenna users by generating a single beam with common data per group. The aim is to minimize the sum power of BSs while satisfying user-specific SINR targets. We propose centralized and distributed multicast beamforming algorithms for multi-cell multigroup systems. The NP-hard multicast problem is tackled by approximating it as a convex problem using the standard semidefinite relaxation method. The resulting semidefinite program (SDP) can be solved via centralized processing if global channel knowledge is available. To allow a distributed implementation, the primal decomposition method is used to turn the SDP into two optimization levels. The higher level is in charge of optimizing inter-cell interference while the lower level optimizes beamformers for given inter-cell interference constraints. The distributed algorithm requires local channel knowledge at each BS and scalar information exchange between BSs. If the solution has unit rank, it is optimal for the original problem. Otherwise, the Gaussian randomization method is used to find a feasible solution. The superiority of the proposed algorithms over conventional schemes is demonstrated via numerical evaluation.
\end{abstract}


\section{Introduction}
\label{sec:intro}
Transmit beamforming (or equivalently precoding) is a signal processing technique that aims at improving the performance of a communication system by efficiently exploiting the spatial domain of a wireless multi-antenna channel.
Advanced multi-antenna beamforming techniques can increase spectral efficiency significantly, if properly designed. However, without proper interference coordination between neighboring cells, inter-cell interference may limit the system performance.
In this respect, coordinated beamforming, where inter-cell interference coordination is involved in the design of multi-antenna techniques, has been recognized as a powerful approach to improve the performance of wireless systems, especially at cell-edge areas \cite{Gesbert-10}.
In coordinated beamforming, each data stream is linearly precoded in the spatial domain and transmitted from a single base station (BS). To control interference, precoded data transmissions are jointly designed
among BSs such that a practical network design target is achieved
while predetermined constraints imposed on users and BSs are satisfied.
The performance of coordinated beamforming schemes rests on the availability of channel state information (CSI) at the BSs.
Coordinated beamforming techniques can be implemented either in a centralized or a decentralized manner.
Centralized algorithms require the knowledge of the channels between all BSs and all users in the system, i.e., global CSI. Distributed approaches rely on the availability of local CSI, i.e., the knowledge of the channels between a BS and all users in the system. Throughout this paper, the acquired global or local CSI is assumed to be perfect.
In general, decentralized schemes are often more practically realizable than the centralized ones due to possibly reduced signaling overhead and lower computational requirements per processing unit.
Coordinated beamforming has been extensively studied for various system design objectives, such as sum power minimization \cite{Rashid-Farrokhi-98a}, minimum SINR maximization \cite{Huang-12} and sum rate maximization \cite{Shi-11}.
In the classical sum power minimization problem \cite{Rashid-Farrokhi-98a}, the goal is to minimize the sum power of the BSs while satisfying user-specific SINR targets.
This system design objective is of practical interest for wireless applications which have stringent data rate and delay constraints.
In the literature, centralized and distributed beamforming algorithms have been proposed in \cite{Rashid-Farrokhi-98a,Bengtsson-99,Bengtsson-01,Wiesel-06} and \cite{Rashid-Farrokhi-98a,Dahrouj-10,Tolli-11,Pennanen-11,Pennanen-14a}, respectively. These algorithms are either based on standard convex optimization techniques or exploitation of uplink-downlink duality.

In the hitherto presented literature, independent data is addressed to each user. This transmission strategy is known as unicast beamforming.
When a symbol is addressed to more than one user, however, a more elaborate multicasting problem arises.
Physical layer multicasting has the potential to efficiently address the nature of future traffic demands, e.g., to support demanding video broadcasting applications.
A physical layer multicasting problem was originally proposed in \cite{Sidiropoulos-06}, proven NP-hard and accurately approximated by the semidefinite relaxation (SDR) and Gaussian randomization techniques.
In \cite{Karipidis-08}, a unified framework was derived for physical layer multigroup multicasting, where independent sets of common data are transmitted to different interfering groups of users.
Therein, the sum power minimization and the minimum SINR maximization problems, also known as the Quality of Service (QoS) and the max-min fair problems, were formulated, proven NP-hard and accurately approximated for a multicast multigroup system with a sum power constraint.
In \cite{Christopoulos-14,Christopoulos-14b}, a consolidated solution was derived for the weighted max-min fair multigroup multicast beamforming under per-antenna power constraints.
This work was extended for the sum rate maximization problem in \cite{Christopoulos-14c}.
In \cite{Xiang-13}, a distributed algorithm was proposed for a multi-cell multicast system with a single group per cell.
Both QoS and max-min fair problems were studied. The energy efficiency maximization problem was recently considered in \cite{He-15} for a multi-cell multigroup case, and a centralized algorithm was derived.
In the literature, however, there is a lack of generic centralized and distributed algorithms for the sum power minimization problem in a multi-cell multicast system with multiple groups per cell.

In the present contribution, in contrast to existing works, centralized and distributed beamforming designs are proposed for a multi-cell multigroup multicast system. The target is to minimize the total transmission power of the system while providing the guaranteed minimum SINRs for active users. This non-convex problem is first approximated as a convex one via the SDR. The resulting semidefinite program (SDP) can be efficiently solved via centralized processing, requiring global channel knowledge.
In order to obtain a distributed implementation, the primal decomposition method is used to reformulate the one-level SDP into two optimization levels. In the higher level, upper bounding constraints for inter-cell interference powers are optimized while the lower level is in charge of optimizing the beamformers for a given set of interference power constraints. Distributed processing requires only local CSI at each BS and the exchange of scalar information with other BSs via low-rate backhaul links.
With rank-one solution, the SDR is optimal for the original problem.
Otherwise, the Gaussian randomization method is utilized to provide a sub-optimal, but feasible, beamforming solution. The effectiveness of the proposed centralized and distributed beamforming schemes is demonstrated via numerical examples.

The paper is organized as follows. In Section \ref{sec:System_Model}, the multi-cell multicast system is introduced, and the corresponding sum power minimization problem is formulated. In sections \ref{sec:CentralizedAlg} and \ref{sec:DecentralizedAlg}, the centralized and distributed beamforming algorithms are derived, respectively. The performance of the proposed algorithms are examined in Section \ref{sec:SimulationResults} via numerical examples. Finally, conclusions are drawn in Section \ref{sec:Conclusion}.
The following notation is used.
Bold face lower case and upper case characters denote column vectors and matrices, respectively. The operators $\left(\cdot\right)\herm$ and $\mathrm{Tr}(\cdot)$ correspond to the conjugate transpose and the trace of a matrix.
${\mathbb R}_{++}^{N}$ denotes the set of $N$-dimensional positive real vectors, while ${\mathbb C}^{M}$ represents the set of $M$-dimensional complex vectors.

\section{System Model and problem formulation}
\label{sec:System_Model}
Consider a multi-cell multigroup multicasting system with $B$ BSs, $G$ groups and $U$ users. The corresponding sets of BSs, groups and users are denoted by $\mathcal{B}=\{1,\ldots,B\}$, $\mathcal{G}=\{1,\ldots,G\}$ and $\mathcal{U}=\{1,\ldots,U\}$, respectively. Each BS is equipped with $A$ transmit antennas, whereas each user has only one receive antenna.
An independent data stream is transmitted to each group of users from a single serving BS.
Thus, there exists inter-group interference between the groups of a serving BS (i.e., intra-cell interference) and inter-group interference between the groups that belong to the different BSs (i.e., inter-cell interference).
The set of groups served by BS $b$ is given by $\mathcal{G}_b$. The number of groups in set $\mathcal{G}_b$ is denoted by $G_b$.
The set of users in group $g$ is denoted by $\mathcal{U}_g$, and the corresponding number of users is given by $U_g$.
Since each user belongs to only one group, the sets of users belonging to different groups are disjoint, i.e., $\mathcal{U}_i \cap \mathcal{U}_j = \oslash$, $\forall i,j \in \mathcal{G}, i \neq j$.
The received signal at user $u$ is given by
\begin{eqnarray} \label{eq:RxSignal}
{y}_{u} &=& \overbrace{{\vec h}_{b,u}\herm {\vec w}_{g} {s}_{g}}^\text{desired signal} + \overbrace{\sum\limits_{i \in \mathcal{G}_{b} \setminus \{g\}} {\vec h}_{b,u}\herm {\vec w}_{i} {s}_{i}}^\text{intra-cell interference} \nonumber \\
&& + \underbrace{\sum\limits_{j \in \mathcal{B} \setminus \{b\}} \sum\limits_{k \in \mathcal{G}_j} {\vec h}_{j,u}\herm {\vec w}_{k} {s}_{k}}_\text{inter-cell interference} + {n}_{u}, \nonumber \\
&& \forall b \in \mathcal{B}, \forall g \in \mathcal{G}_b, \forall u \in \mathcal{U}_g
\end{eqnarray}
where ${\vec h}_{b,u} \in \C^{A}$ is the channel vector from BS $b$ to user $u$, ${\vec w}_{g} \in \C^{A}$ is the transmit beamforming vector of group $g$, ${s}_{g} \in \C$ is the corresponding normalized data symbol and ${n}_{u} \ \sim \mathcal{C} \mathcal{N} (0, \sigma_{u}^{2})$ is the complex white Gaussian noise sample with zero mean and variance $\sigma_{u}^{2}$.

The system optimization objective is to minimize the total transmission power of all BSs while guaranteeing minimum SINR target for each active user. The mathematical expression of the problem is given by
\begin{equation} \label{eq:SPMinMulticast}
\begin{array}{ll}
\displaystyle \underset{\{{\vec w}_{g}\}_{g \in \mathcal{G}}}{\mathrm{min.}}  &  \displaystyle  \sum\limits_{g \in \mathcal{G}} {\rm Tr} \left({\vec w}_{g} {\vec w}_{g}\herm \right)\\
{\mathrm{s.\ t.}}
& \displaystyle \frac{|{\vec h}_{b,u}\herm {\vec w}_{g}|^{2}}{{\sigma_{u}^{2} + \sum\limits_{j \in \mathcal{B}} \sum\limits_{k \in \mathcal{G}_j \setminus \{g\}} |{\vec h}_{j,u}\herm {\vec w}_{k}|^{2}}} \geq \gamma_{u}, \\
& \forall b \in \mathcal{B}, \forall g \in \mathcal{G}_b, \forall u \in \mathcal{U}_g
\end{array}
\end{equation}
where $\gamma_{u}$ is the minimum SINR target for user $u$.
Problem \eqref{eq:SPMinMulticast} can be infeasible in some channel conditions and system settings, e.g., the predetermined SINR targets and/or the number of active users are too high.
In general, it is the duty of admission control to handle infeasible cases by relaxing the system requirements, i.e., by decreasing the SINR targets and/or reducing the number of users \cite{Stridh-06}.
Feasibility was discussed for unicast and multicast beamforming systems in \cite{Wiesel-06} and \cite{Xiang-13}, respectively. In the rest of this paper, \eqref{eq:SPMinMulticast} is assumed to be feasible.
Problem \eqref{eq:SPMinMulticast} is non-convex and NP-hard since it is a more generic version of an NP-hard single-cell multicast problem \cite{Sidiropoulos-06}. Thus, \eqref{eq:SPMinMulticast} cannot be solved in its current form.

\section{Centralized beamforming design}
\label{sec:CentralizedAlg}
Problem \eqref{eq:SPMinMulticast} can be approximated as a convex problem, which can be efficiently solved.
In this respect, the SDR method is applied by replacing ${\vec w}_{g}{\vec w}_{g}\herm$ with a semidefinite matrix ${\vec W}_{g}$, $\forall g \in \mathcal{G}$.
The relaxation lets the rank of ${\vec W}_{g}$ be arbitrary.
The resulting convex SDP is expressed as
\begin{equation} \label{eq:SPMinMulticastApproximated}
\begin{array}{ll}
\displaystyle \underset{\{{\vec W}_{g}\}_{g \in \mathcal{G}}}{\mathrm{min.}}  &  \displaystyle  \sum\limits_{g \in \mathcal{G}} {\rm Tr} \left({\vec W}_{g}\right)\\
{\mathrm{s.\ t.}}
& \displaystyle \frac{{\rm Tr} \left({\vec H}_{b,u}{\vec W}_{g}\right)}{{\sigma_{u}^{2} + \sum\limits_{j \in \mathcal{B}} \sum\limits_{k \in \mathcal{G}_j \setminus \{g\}} {\rm Tr} \left({\vec H}_{j,u} {\vec W}_{k} \right)}} \geq \gamma_{u}, \\
& \forall b \in \mathcal{B}, \forall g \in \mathcal{G}_b, \forall u \in \mathcal{U}_g \\
& {\vec W}_{g} \succeq 0, \forall g \in \mathcal{G}
\end{array}
\end{equation}
where ${\vec H}_{b,u}={\vec h}_{b,u}{\vec h}_{b,u}\herm$. Problem \eqref{eq:SPMinMulticastApproximated} can be solved in a centralized way if global CSI is available at a central controlling unit or at each BS.
An optimal solution of \eqref{eq:SPMinMulticastApproximated} is not necessarily optimal for the original non-convex problem \eqref{eq:SPMinMulticast}. If the solution is rank-one, i.e., all the optimal transmit covariance matrices $\{{\vec W}_{g}^{*}\}_{g \in \mathcal{G}}$ have unit ranks, then the solution is also optimal for the original problem.
In this case, the optimal beamformers $\{{\vec w}_{g}^{*}\}_{g \in \mathcal{G}}$ can be extracted from $\{{\vec W}_{g}^{*}\}_{g \in \mathcal{G}}$ by using the eigenvalue decomposition. The resulting beamformers are given by ${\vec w}_{g}^{*}=\sqrt{\lambda_{g}}{\vec u}_{g}$, $\forall g \in \mathcal{G}$,
where $\lambda_{g}$ and ${\vec u}_{g}$ are the principal eigenvalue and eigenvector of ${\vec W}_{g}^{*}$.

For specific optimization problems, the SDR provides optimum solutions.
The most prominent example of this case is the optimal unicast beamforming solution in \cite{Bengtsson-01}.
Nevertheless, due to the NP-hardness of the multicast problem, the relaxed problems do not necessarily yield unit rank matrices. Consequently, one can  apply a rank-one approximation over the higher rank solution.
The Gaussian randomization method is reported to give the highest accuracy in the multicast beamforming case \cite{Luo-10}.
Let the symmetric positive semidefinite matrices $\{{\vec W}_{g}^{*}\}_{g \in \mathcal{G}}$ constitute a solution of the relaxed problem.
Then, a candidate rank-one beamforming solution to the original problem can be generated as a complex Gaussian vector with zero mean and covariance equal to ${\vec W}_{g}^{*}$, i.e.  $\hat{{\vec w}}_g \ \sim \mathcal{C} \mathcal{N}(0, {\vec W}_{g}^{*} )$, $\forall g \in \mathcal{G}$.
Next, an intermediate step is required between generating a Gaussian instance with the statistics obtained from the relaxed solution and creating a feasible candidate instance of the original problem since the feasibility of the original problem is not guaranteed.
In this respect, an additional power minimization problem needs to be solved. For a given set of candidate beamformers $\{\hat{{\vec w}}_{g}\}_{g \in \mathcal{G}}$, the transmission powers $\{p_g\}_{g \in \mathcal{G}}$ are minimized while satisfying the user-specific SINR targets $\{\gamma_u\}_{u \in \mathcal{U}}$. The resulting linear program (LP) is given by
\begin{equation} \label{eq:GR_PowOpt_Centr}
\begin{array}{ll}
\displaystyle \underset{\{p_{g}\}_{g \in \mathcal{G}}}{\mathrm{min.}}  &  \displaystyle  \sum\limits_{g \in \mathcal{G}} p_{g}\\
{\mathrm{s.\ t.}}
& \displaystyle \frac{p_g \left|{\vec h}_{b,u}\hat{{\vec w}}_{g}\right|^{2}}{{\sigma_{u}^{2} + \sum\limits_{j \in \mathcal{B}} \sum\limits_{k \in \mathcal{G}_j \setminus \{g\}} p_k \left|{\vec h}_{j,u}\hat{{\vec w}}_{k}\right|^{2}}} \geq \gamma_{u}, \\
& \forall b \in \mathcal{B}, \forall g \in \mathcal{G}_b, \forall u \in \mathcal{U}_g.
\end{array}
\end{equation}
By solving \eqref{eq:GR_PowOpt_Centr}, a set of beamformers is defined by ${\vec w}_{g}=\sqrt{p_{g}^{*}} \hat{{\vec w}}_{g}$, $\forall g \in \mathcal{G}$, where $p_{g}^{*}$ is the optimal power associated with fixed candidate beamformer $\hat{{\vec w}}_{g}$.
The beamformers $\{{\vec w}_{g}\}_{g \in \mathcal{G}}$ are sub-optimal, but feasible, for the original problem.
Finally, after generating a predetermined number of candidate solutions, the one that yields the lowest objective value of the original problem is chosen.
The accuracy of this approximate solution is measured by the distance of the approximate objective value and the optimal value of the relaxed problem. This accuracy increases with the increasing number of Gaussian randomizations.
The proposed centralized multicast approach is summarized in {\it Algorithm~\ref{alg:SPMinMulticastAlgCentr}}.
With global CSI, {\it Algorithm~\ref{alg:SPMinMulticastAlgCentr}} is performed at a central controlling unit or at BS $b$, for all $b$ in parallel.

\begin{algorithm} [tbp!]
\caption{Centralized multicast beamforming}
\label{alg:SPMinMulticastAlgCentr}
\begin{algorithmic}[1]
\STATE Compute optimal transmit covariance matrices $\{{\vec W}_{g}^{*}\}_{g \in \mathcal{G}}$ by solving the relaxed problem as an SDP \eqref{eq:SPMinMulticastApproximated}.
\STATE Check whether the ranks of $\{{\vec W}_{g}^{*}\}_{g \in \mathcal{G}}$ are all one or not. If the ranks are one, apply eigenvalue decomposition for $\{{\vec W}_{g}^{*}\}_{g \in \mathcal{G}}$ to find optimal beamformers $\{{\vec w}_{g}^{*}\}_{g \in \mathcal{G}}$ for the original problem. Otherwise, apply Gaussian randomization with power optimization \eqref{eq:GR_PowOpt_Centr} to find feasible, but sub-optimal, beamformers $\{{\vec w}_{g}\}_{g \in \mathcal{G}}$.
\end{algorithmic}
\end{algorithm}

\section{Distributed beamforming design}
\label{sec:DecentralizedAlg}
In this section, a primal decomposition-based distributed beamforming approach is proposed. Primal decomposition method can be used to facilitate distributed implementation since it decouples the problem at each iteration.
This method can be applied to an optimization problem which has such coupling constraints that by fixing them, the problem decouples.
In the following, we first reformulate the centralized SDR problem, and then apply primal decomposition. By using primal decomposition, the one-level optimization problem is divided into two levels, i.e., the lower level subproblems and the higher level master problem. The solution method for this two-level optimization is derived. The conditions for the optimality of the obtained solution with respect to the original problem are described.
Gaussian randomization method is presented in case the solution is not optimal for the original problem. Finally, the distributed approach is summarized by a step-by-step algorithm, and its practical properties are discussed.

\subsection{Reformulation of the centralized relaxed problem}
\label{sec:ReformulatedProblem}
In order to apply primal decomposition, \eqref{eq:SPMinMulticastApproximated} needs to be reformulated by adding auxiliary variables. In this respect, we separate interference power to intra-cell and inter-cell terms, and add auxiliary variables to denote the inter-cell interference terms.
Now, the coupling is transferred from beamformers to inter-cell interference variables.
The reformulated problem is expressed as
\begin{equation} \label{eq:SPMinMulticastApproximatedReform}
\begin{array}{ll}
\displaystyle \underset{\{{\vec W}_{g}\}_{g \in \mathcal{G}}, {\boldsymbol \theta} }{\mathrm{min.}}  &  \displaystyle  \sum\limits_{g \in \mathcal{G}} {\rm Tr} \left({\vec W}_{g}\right)\\
{\mathrm{s.\ t.}}
& \displaystyle \hspace{-0.5cm} \frac{{\rm Tr} \left({\vec H}_{b,u}{\vec W}_{g}\right)}{{\sigma_{u}^{2} + \sum\limits_{j \in \mathcal{B} \setminus \{b\}} \theta_{j,u} + \sum\limits_{k \in \mathcal{G}_{b} \setminus \{g\}} {\rm Tr} \left({\vec H}_{b,u} {\vec W}_{k} \right)}} \geq \gamma_{u}, \\
& \hspace{-0.5cm} \forall b \in \mathcal{B}, \forall g \in \mathcal{G}_b, \forall u \in \mathcal{U}_g \\
& \hspace{-0.5cm} \sum\limits_{i \in \mathcal{G}_{b}} {\rm Tr} \left({\vec H}_{b,u} {\vec W}_{i} \right) \leq \theta_{b,u}, \forall b \in \mathcal{B}, \forall u \in \mathcal{U} \setminus \mathcal{U}_b \\
& \hspace{-0.5cm} {\vec W}_{g} \succeq 0, \forall g \in \mathcal{G}_b, \forall b \in \mathcal{B}
\end{array}
\end{equation}
where $\theta_{b,u}$ is the inter-cell interference from BS $b$ to user $u$ and the vector ${\boldsymbol \theta}$ consists of all inter-cell interference variables. Since the inequality constraints are met with equality at the optimal solution, \eqref{eq:SPMinMulticastApproximatedReform} yields the same solution than \eqref{eq:SPMinMulticastApproximated}.

\subsection{Two-level optimization via primal decomposition}
\label{sec:TwoLevelProblem}
By applying primal decomposition, \eqref{eq:SPMinMulticastApproximatedReform} is divided into BS-specific subproblems for beamforming design with fixed inter-cell interference levels, and a network wide master problem in charge of optimizing the interference levels.
The resulting subproblem for BS $b$ is given by
\begin{equation} \label{eq:SPMinMulticastSubproblem}
\begin{array}{ll}
\displaystyle \underset{\{{\vec W}_{g}\}_{g \in \mathcal{G}_b}}{\mathrm{min.}}  &  \displaystyle  \sum\limits_{g \in \mathcal{G}_b} {\rm Tr} \left({\vec W}_{g}\right)\\
{\mathrm{s.\ t.}}
& \displaystyle \hspace{-0.4cm} \frac{{\rm Tr} \left({\vec H}_{b,u}{\vec W}_{g}\right)}{{\sigma_{u}^{2} + \sum\limits_{j \in \mathcal{B} \setminus \{b\}} \theta_{j,u} + \sum\limits_{k \in \mathcal{G}_{b} \setminus \{g\}} {\rm Tr} \left({\vec H}_{b,u} {\vec W}_{k} \right)}} \geq \gamma_{u}, \\
& \hspace{-0.4cm} \forall g \in \mathcal{G}_b, \forall u \in \mathcal{U}_g \\
& \hspace{-0.4cm} \sum\limits_{i \in \mathcal{G}_{b}} {\rm Tr} \left({\vec H}_{b,u} {\vec W}_{i} \right) \leq \theta_{b,u}, \in \mathcal{U} \setminus \mathcal{U}_b \\
& \hspace{-0.4cm} {\vec W}_{g} \succeq 0, \forall g \in \mathcal{G}_b,
\end{array}
\end{equation}
Problem \eqref{eq:SPMinMulticastSubproblem} can be optimally solved since it is an SDP.
The master problem is given by
\begin{equation} \label{eq:SPMinMulticastMasterProblem}
\begin{array}{cl}  \underset{\{{\boldsymbol \theta}_{b}\}_{b \in \mathcal{B}}}{\mathrm{min.}}  &  \sum\limits_{b \in \mathcal{B}} f^{\star}_{b} ({\boldsymbol \theta}_{b}) \\
{\mathrm{s.\ t.}} & {\boldsymbol \theta}_{b} \in \R^{L}_{++}, \forall b \in \mathcal{B} \\
\end{array}
\end{equation}
where $f^{\star}_{b} ({\boldsymbol \theta}_{b})$ denotes the optimal objective value of \eqref{eq:SPMinMulticastSubproblem} for given ${\boldsymbol \theta}_{b}$. The vector ${\boldsymbol \theta}_{b}$ with length $L$ is composed of BS $b$ specific inter-cell interference terms.
The master problem \eqref{eq:SPMinMulticastMasterProblem} can be solved for the inter-cell interference variables $\{\theta_{b,u}\}_{b \in \mathcal{B}, u \in \mathcal{U} \setminus \mathcal{U}_b}$ by using the projected subgradient method
\begin{eqnarray} \label{eq:SubgradientMethod}
\theta_{b,u}^{(r+1)} & = & \mathcal{P} \left \{ \theta_{b,u}^{(r)} - \sigma^{(r)} s_{b,u}^{(r)} \right \}, b \in \mathcal{B}, u \in \mathcal{U} \setminus \mathcal{U}_b
\end{eqnarray}
where $\mathcal{P}$ is the projection onto a positive orthant, $r$ is the iteration index, $\sigma^{(r)}$ is the step-size and $s_{b,u}^{(r)}$ is the subgradient of \eqref{eq:SPMinMulticastMasterProblem} at point $\theta_{b,u}^{(r)}$. Due to the convexity of problem \eqref{eq:SPMinMulticastApproximatedReform}, the subgradient $s_{b,u}^{(r)}$ can be defined via the dual problem of \eqref{eq:SPMinMulticastApproximatedReform} by using similar derivation as in \cite{Pennanen-14a}. The resulting subgradient at point $\theta_{b,u}^{(r)}$ is given by $s_{b,u}^{(r)} = \lambda_{b,u}^{(r)} - \mu_{j,u}^{(r)}$,
where $\lambda_{b,u}^{(r)}$ is the dual variable associated with $\theta_{b,u}^{(r)}$ in the SINR constraint of user $u$ at its serving BS $b$ (i.e., in subproblem $b$) and $\mu_{j,u}^{(r)}$ is the dual variable associated with $\theta_{b,u}^{(r)}$ in the inter-cell interference constraint of user $u$ at the interfering BS $j$ (i.e., in subproblem $j$).
Since \eqref{eq:SPMinMulticastSubproblem} is convex, the optimal dual variables can be obtained as side information (i.e., a certificate for optimality) by solving \eqref{eq:SPMinMulticastSubproblem} using standard SDP solvers. An alternative and explicit way to find the dual variables is to formulate and solve the dual problem of \eqref{eq:SPMinMulticastSubproblem}.

The master problem can be optimally solved if the step-size of the projected subgradient method is properly chosen \cite{Palomar-06}.
If local CSI is available and a small amount of information exchange is allowed between the BSs, a distributed implementation is possible, es explained in Section \ref{sec:SummaryAlg}.
If all the optimal covariance matrices $\{{\vec W}_{g}^{*}\}_{g \in \mathcal{G}_b, b \in \mathcal{B}}$ have unit ranks, then this solution is also optimal for the original problem \eqref{eq:SPMinMulticast}. In this case, the optimal beamformers $\{{\vec w}_{g}^{*}\}_{g \in \mathcal{G}_b, b \in \mathcal{B}}$ are obtained from $\{{\vec W}_{g}^{*}\}_{g \in \mathcal{G}_b, b \in \mathcal{B}}$ by applying the eigenvalue decomposition, i.e., ${\vec w}_{g}^{*}=\sqrt{\lambda_{g}}{\vec u}_{g}$, $\forall g \in \mathcal{G}_b$, $\forall b \in \mathcal{B}$.

\subsection{Gaussian randomization}
\label{sec:GR}
If at least one of $\{{\vec W}_{g}^{*}\}_{g \in \mathcal{G}_b, b \in \mathcal{B}}$ has a rank higher than one, the solution of the SDR problem is not optimal for the original problem \eqref{eq:SPMinMulticast}.
In this case, feasible, but sub-optimal, rank-one beamformers can be found via Gaussian randomization method.
A candidate beamforming solution $\hat{{\vec w}}_{g}$, $\forall g \in \mathcal{G}_b$, $\forall b \in \mathcal{B}$, is generated as a Gaussian random variable with zero mean and covariance ${\vec W}_{g}^{*}$.
Since the candidate beamformers may not be feasible to the original problem as such, an additional power optimization problem needs to be solved at each BS. At BS $b$, powers $\{p_g\}_{g \in \mathcal{G}_b}$ are optimized for a given set of fixed candidate beamformers $\{\hat{{\vec w}}_g\}_{g \in \mathcal{G}_b}$ while the SINR targets $\{\gamma_u\}_{u \in \mathcal{U}_b}$ need to be satisfied.
This problem can be expressed as the following LP
\begin{equation} \label{eq:GR_PowOpt}
\begin{array}{ll}
\displaystyle \underset{\{p_{g}\}_{g \in \mathcal{G}_b}}{\mathrm{min.}}  &  \displaystyle  \sum\limits_{g \in \mathcal{G}_b} p_{g}\\
{\mathrm{s.\ t.}}
& \displaystyle \frac{p_g \left|{\vec h}_{b,u}\hat{{\vec w}}_{g}\right|^{2}}{{\sigma_{u}^{2} + \sum\limits_{j \in \mathcal{B} \setminus \{b\}} \theta_{j,u} + \sum\limits_{k \in \mathcal{G}_{b} \setminus \{g\}} p_k \left|{\vec h}_{b,u}\hat{{\vec w}}_{k}\right|^{2}}} \geq \gamma_{u}, \\
& \forall g \in \mathcal{G}_b, \forall u \in \mathcal{U}_g \\
& \sum\limits_{i \in \mathcal{G}_{b}} p_i \left|{\vec h}_{b,u}\hat{{\vec w}}_{i}\right|^{2} \leq \theta_{b,u}, \in \mathcal{U} \setminus \mathcal{U}_b. \\
\end{array}
\end{equation}
After solving \eqref{eq:GR_PowOpt}, BS $b$ can define its beamformers by ${\vec w}_{g}=\sqrt{p_{g}^{*}} \hat{{\vec w}}_{g}$, $\forall g \in \mathcal{G}_b$, where $p_{g}^{*}$ is the optimal power associated with the candidate beamformer $\hat{{\vec w}}_{g}$.
The resulting beamformers are sub-optimal for the original problem.
After generating a predefined number of candidate solutions, the one that gives the lowest objective value of the original problem is selected.
Solving \eqref{eq:GR_PowOpt} does not require any information exchange between the BSs since the inter-cell interference variables are fixed while only the powers are optimized. The fixed values are taken from the optimal solution of \eqref{eq:SPMinMulticastMasterProblem}.
An alternative problem formulation is possible where both powers and inter-cell interference variables are optimized simultaneously with the aid of iterative primal decomposition method. Solving this problem requires scalar information exchange among the BSs via backhaul.

\subsection{Distributed implementation}
\label{sec:SummaryAlg}
The distributed implementation of the beamforming design is enabled if each BS acquires local CSI and scalar information exchange between the BSs is allowed via low-rate backhaul links.
More precisely, the subproblem $b$ in \eqref{eq:SPMinMulticastSubproblem} and the corresponding part of the master problem in \eqref{eq:SPMinMulticastMasterProblem}, i.e., the update of ${\boldsymbol \theta}_{b}$, are solved independently at BS $b$, for all $b \in \mathcal{B}$ in parallel.
At subgradient iteration $r$, the backhaul information exchange is performed by BS $b$ as follows. BS $b$ signals the dual variables associated with the SINR constraints, i.e., $\{\lambda_{b,u}\}_{u \in \mathcal{U}_{b}}$, to all the interfering BSs. Whereas the dual variables associated with the inter-cell interference constraints, i.e., $\{\mu_{b,u}\}_{u \in \mathcal{U} \setminus \mathcal{U}_{b}}$, are signaled to the BS of which user is being interfered by BS $b$.
Assuming a fully connected network and an equal number of users at each cell (i.e., $U_b=U/B$, $\forall b \in \mathcal{B}$), the total amount of the required backhaul signaling at each subgradient iteration $r$ is the sum of the real-valued terms exchanged between the coupled BS pairs. Thus, the total number of exchanged scalar values per iteration is given by $2B(B-1)(U/B)$.
After solving the SDR problem \eqref{eq:SPMinMulticastApproximatedReform} via iterative distributed optimization, each BS needs to know if the covariance matrices of other BSs are all rank-one. This is easily handled in a distributed manner by each BS sending one-bit feedback to other BSs. If the Gaussian randomization procedure needs to be used, extra backhaul signaling is required. More precisely, the BS-specific powers for each Gaussian randomization instance need to be shared among other BSs in order to select the best one in a distributed manner.
The overall distributed approach is summarized in {\it Algorithm~\ref{alg:SPMinMulticastAlg}}. {\it Algorithm~\ref{alg:SPMinMulticastAlg}} is performed at BS $b$, for all $b$ in parallel.

\begin{algorithm} [tbp!]
\caption{Distributed multicast beamforming}
\label{alg:SPMinMulticastAlg}
\begin{algorithmic}[1]
\STATE Set $r=0$. Initialize inter-cell interference powers $\boldsymbol{\theta}_{b}^{(0)}$.
\REPEAT
\STATE Compute optimal transmit covariance matrices $\{{\vec W}_{g}^{*}\}_{g \in \mathcal{G}_b}$ and dual variables $\{{\lambda}_{b,u}\}_{u \in \mathcal{U}_b}$, $\{{\mu}_{b,u}\}_{u \in \mathcal{U} \setminus \mathcal{U}_b}$ by solving the relaxed subproblem $b$ as an SDP \eqref{eq:SPMinMulticastSubproblem}.
\STATE Communicate dual variables $\{{\lambda}_{b,u}\}_{u \in \mathcal{U}_b}$, $\{{\mu}_{b,u}\}_{u \in \mathcal{U} \setminus \mathcal{U}_b}$ to the coupled BSs via backhaul.
\STATE Update inter-cell interference variables $\boldsymbol {\theta}_{b}^{(r+1)}$ via projected subgradient method \eqref{eq:SubgradientMethod}.
\STATE Set $r=r+1$.
\UNTIL{desired level of convergence}
\STATE Check whether the ranks of $\{{\vec W}_{g}^{*}\}_{g \in \mathcal{G}_b}$ are all one or not. Share this one-bit information among other BSs via backhaul. If the ranks are one for all $g \in \mathcal{G}_b, b \in \mathcal{B}$, apply eigenvalue decomposition for $\{{\vec W}_{g}^{*}\}_{g \in \mathcal{G}_b}$ to find optimal beamformers $\{{\vec w}_{g}^{*}\}_{g \in \mathcal{G}_b}$ for the original problem. Otherwise, apply Gaussian randomization with power optimization \eqref{eq:GR_PowOpt} to find feasible, but sub-optimal, beamformers $\{{\vec w}_{g}\}_{g \in \mathcal{G}_b}$.
\end{algorithmic}
\end{algorithm}

\subsection{Practical considerations}
\label{sec:PracticalConsiderations}
To acquire optimal performance, {\it Algorithm~\ref{alg:SPMinMulticastAlg}} needs to be run until convergence, and provided that the obtained covariance matrices are all rank-one.
However, this is somewhat impractical since the more iterations are run, the higher the signaling/computational load and the longer the caused delay.
In this respect, {\it Algorithm~\ref{alg:SPMinMulticastAlg}} naturally lends itself to a practical design where it can be stopped after a limited number of iterations to reduce delay and signaling load. Since the inter-cell interference levels are fixed at each iteration, feasible beamformers can be computed via the eigenvalue decomposition or the Gaussian randomization procedure, depending on the rank properties of the covariance matrices.
Limiting the number of iterations comes at the cost of increased sum power.

In Table \ref{tab:SignalingDistributed}, the backhaul signaling overhead of the centralized and distributed algorithms are compared under different system settings.
In the centralized algorithm, it is assumed that each BS exchanges its local CSI with all other BSs via backhaul links. Thus, global CSI is made available for each BS.
Assuming equal number of users at each cell, the total backhaul signaling load in terms of scalar-valued channel coefficients in the centralized system is given by $2AU(B-1)B$.
Here, one complex channel coefficient is considered as two real-valued coefficients.
For the distributed algorithm, the total backhaul signaling load is presented per subgradient iteration.
In Table \ref{tab:SignalingDistributed}, the values inside the brackets denote the percentage of the signaling load required per distributed iteration, compared with the overall signaling load required by the centralized algorithm.
One can see that the distributed algorithm requires notably less amount of backhaul signaling per iteration compared to the centralized approach. The difference gets greater with the increasing network size.
In conclusion, backhaul signaling overhead can be significantly reduced by limiting the number of iterations.

The distributed approach allows some special case designs where the number of optimization variables is reduced, leading to a lower computational load and even a further decreased signaling overhead. These special case designs come at the cost of somewhat decreased performance. Some of the possible special cases are presented below:
\begin{itemize}
\item Common interference constraint: $\theta_{b,u}=\theta, \forall b \in \mathcal{B}, \forall u \in \mathcal{U} \setminus {\mathcal{U}}_b$.
\item Fixed interference constraints: $\theta_{b,u}=c_{b,u}$, $\forall b \in \mathcal{B}, \forall u \in \mathcal{U} \setminus {\mathcal{U}}_b$, where $c_{b,u}$ is a predefined constant. Does not require any backhaul signaling.
\item Inter-cell interference nulling, i.e., $\theta_{b,u}=0$, $\forall b \in \mathcal{B}, \forall u \in \mathcal{U} \setminus {\mathcal{U}}_b$.
    Does not require any backhaul signaling.
\end{itemize}

\vspace{0.1cm}
\begin{table}[h!]
\centering
\caption{Total backhaul signaling load (per iteration).}\label{tab:SignalingDistributed}\vspace{2pt}
\begin{tabular}{c|c|c}
& Centralized & Distributed \\ \hline
$\{B,U,A\}=\{2,8,8\}$ & 256 & 16 (6.3$\%$) \\ \hline
$\{B,U,A\}=\{3,12,12\}$ & 1728 & 48 (2.8$\%$) \\ \hline
$\{B,U,A\}=\{4,16,16\}$ & 6144 & 96 (1.6$\%$) \\ \hline
\end{tabular}
\end{table}

\section{Simulation results}
\label{sec:SimulationResults}
In this section, the performance of the proposed centralized and distributed algorithms is evaluated via numerical examples.
First, the convergence behavior of the distributed algorithm is examined, and its performance after limited number of iterations is compared to the centralized approach.
Then, the use of coordinated multicast beamforming (i.e., the proposed centralized algorithm) is justified by showing its superiority over conventional transmission schemes.
The performance of the centralized algorithm is also studied against the lower bound solution under different system settings. Finally, the tightness of the SDR method and the properties of the higher rank solutions are also examined.
The used simulation model consists of $B$  BSs, each of which is equipped with $A$ transmit antennas and serves $G$ groups of $U$ single antenna users. The number of user per each group is given by $U/G$.
In the  figures hereafter, the main system parameters are given by $\{B,G,U,A\}$.
We assume frequency-flat Rayleigh fading channel conditions with uncorrelated channel coefficients between antennas.
The SINR constraints are set equal for all users, i.e., $\gamma_u=\gamma$, $\forall u \in \mathcal{U}$. The simulation results are achieved by averaging over $100$ channel realizations. In the case of higher rank covariance matrices, $100$ Gaussian randomizations are generated.

\begin{figure}[tbp!]
  \centering
  \includegraphics[width=6cm]{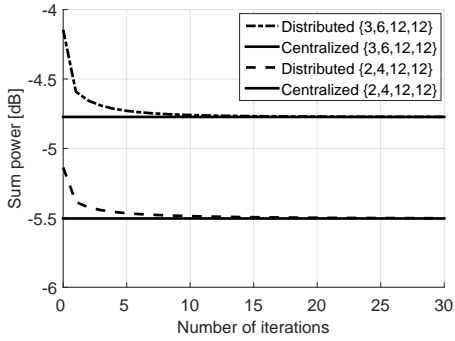}
  \vspace{-0.3cm}
 \caption{Convergence behavior of distributed algorithm.}
\label{fig:Convergence}
\end{figure}

\begin{figure}[tbp!]
  \centering
  \includegraphics[width=6cm]{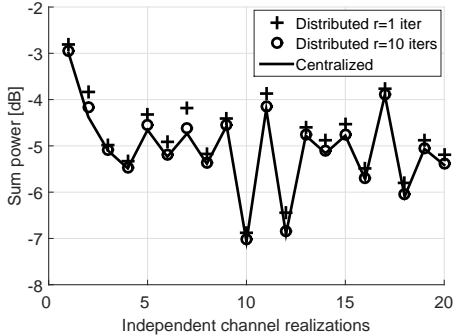}
  \vspace{-0.3cm}
 \caption{Comparison of centralized and distributed algorithms.}
\label{fig:Centr_vs_Distr}
\end{figure}

In Fig.~\ref{fig:Convergence}, the convergence behavior of the distributed algorithm is examined under different system settings.
In this example, the speed of convergence is relatively fast. Especially, the first few iterations improve the performance significantly, and after $10$ iterations the algorithms has almost converged.
In Fig.~\ref{fig:Centr_vs_Distr}, sum power is plotted against independent channel realizations for the centralized and distributed algorithms.
The number of iterations is limited for the distributed algorithm. The main system parameters are given by $\{B,G,U,A\}=\{2,4,8,8\}$. The results demonstrate that the performance of the distributed algorithm with $10$ subgradient iterations is very close to that of the centralized scheme. It can be seen that performance is relatively good even with $1$ iteration.
For Figs.~\ref{fig:Convergence} and~\ref{fig:Centr_vs_Distr}, the SINR target was set to $0$ dB. All the covariance matrices in these results were rank-one.

\begin{figure}[tbp!]
  \centering
  \includegraphics[width=6cm]{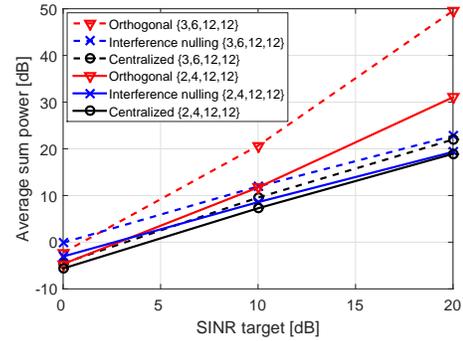}
  \vspace{-0.3cm}
 \caption{Sum power versus SINR target for different transmission schemes.}
\label{fig:PvsSINR}
\end{figure}

In Fig.~\ref{fig:PvsSINR}, average sum power is illustrated against SINR target for various transmission schemes under different system settings.
The following schemes are compared:
\begin{itemize}
\item Single-cell beamforming with orthogonal access (extension of unicast case in \cite{Tolli-09c} to multicast)
\item Coordinated beamforming with inter-cell interference nulling (proposed special case design in Section \ref{sec:PracticalConsiderations})
\item Coordinated beamforming with inter-cell interference optimization (proposed centralized design in Section \ref{sec:CentralizedAlg})
\end{itemize}
In the orthogonal access scheme, each BS uses independent time or frequency slot to optimize the beamformers for its own users leading to an inter-cell interference free communication scenario.
However, the rate target of each user needs to be $B$ times higher as in the non-orthogonal multi-cell case in order to guarantee the same SINR targets. The inter-cell interference nulling scheme forces interference towards other cells' users to be zero via spatial processing.
For simplicity, the results for coordinated beamforming in Fig.~\ref{fig:PvsSINR} were obtained via centralized processing.
However, the same results can be achieved via distributed algorithm if it is let to converge.
The numerical results show that the proposed coordinated beamforming method outperforms the conventional transmission schemes.
Significant performance gains over the interference nulling scheme are witnessed mainly for low and medium SINR targets. The gain diminishes with the increasing SINR target. On the other hand, the superiority against the orthogonal access scheme is greatly emphasized as the SINR target or the number of BSs increases.

\begin{figure}[tbp!]
  \centering
  \includegraphics[width=6cm]{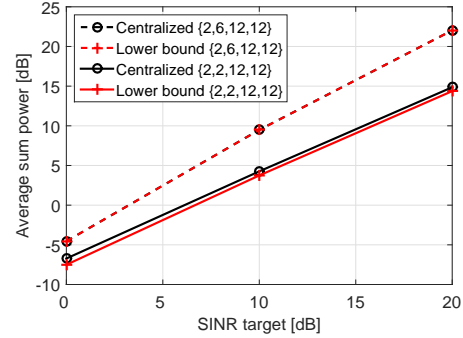}
  \vspace{-0.3cm}
 \caption{Sum power versus SINR target.}
\label{fig:PvsSINR2}
\end{figure}

\begin{figure}[tbp!]
  \centering
  \includegraphics[width=6cm]{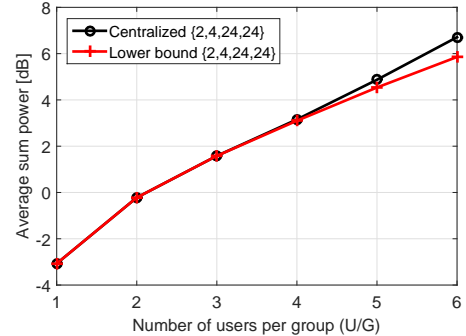}
  \vspace{-0.3cm}
 \caption{Sum power versus the number of users per group.}
\label{fig:PvsUsersPerGroup}
\end{figure}

In Fig.~\ref{fig:PvsSINR2}, the centralized algorithm is compared to the lower bound solution of the relaxed problem. When the solution is rank-one, the SDR is optimal and gives the lower bound. Otherwise, the Gaussian randomization process needs to be applied to get a feasible rank-one solution, which is then compared to the lower bound higher rank solution.
The results imply that the SDR is (usually) optimal when the number of users per group is low (i.e., $U/G=2$) irrespective of the SINR target. If the number of users per group is high (i.e., $U/G=6$), some solutions of the SDR problem have higher rank than one.
Hence, the Gaussian randomization method needs to be used leading to a small gap between the feasible rank-one result and the lower bound solution.
In Fig.~\ref{fig:PvsUsersPerGroup}, the effect of increasing the number of users per group is further studied. Specifically, sum power is presented against the number of users per group.
For low number of users, it seems that the SDR is optimal since it gives the same solution as the lower bound.
However, the performance degrades as the number of users increases, and the gap between the approximation method and the lower bound gets larger. The SINR target was set to $10$ dB.

Table \ref{tab:Rank1} presents the probability that the solution of \eqref{eq:SPMinMulticastApproximated} is rank-one for the increasing number of users per group and for different SINR target values. The results were obtained by averaging over $5000$ channel realizations. The system parameters are given by $\{B,G,U,A\}=\{2,4,4-24,24\}$.
One can see that the probability depends heavily on the number of users per group, while the SINR target has less impact.
More precisely, the probability of rank-one solution decreases as the number of users per group increases. For example, the probability is $100 \%$ for $U/G=1$ and $U/G=2$, while it is less than $25 \%$ for $U/G=6$.
In Table \ref{tab:AveRank}, the average ranks of the higher rank solutions of \eqref{eq:SPMinMulticastApproximated} are illustrated. The parameter setting is identical with Table \ref{tab:Rank1}. The average rank is calculated by summing the ranks of all transmit covariance matrices and dividing it by the number of groups $G$, and then averaging it over $5000$ channel realizations. It can be seen that the average rank slightly increases as the number of users per group increases. Since the dimension of each transmit covariance matrix is 24, the maximum rank could be 24. However, the results demonstrate that the average ranks are relatively low, i.e., always below $1.5$. 

\vspace{0.2cm}
\begin{table}[h!]
\centering
\caption{Probability of rank-one solutions ($\%$).}\label{tab:Rank1}
\begin{tabular}{c|c|c|c|c|c|c}
$U/G$ & 1 & 2 & 3 & 4 & 5 & 6 \\ \hline
$\gamma=0$ dB   & 100 & 100 & 99.78 & 79.0 & 48.4 & 22.8 \\ \hline
$\gamma=10$ dB  & 100 & 100 & 99.86 & 78.7 & 45.6 & 24.2 \\ \hline
$\gamma=20$ dB  & 100 & 100 & 99.98 & 75.2 & 41.3 & 19.7 \\ \hline
\end{tabular}
\vspace{0.2cm}
\end{table}

\vspace{0.2cm}
\begin{table}[h!]
\centering
\caption{Average rank of higher rank solutions.}\label{tab:AveRank}
\begin{tabular}{c|c|c|c|c|c|c}
$U/G$ & 1 & 2 & 3 & 4 & 5 & 6 \\ \hline
$\gamma=0$ dB   & - & - & 1.25 & 1.27 & 1.32 & 1.40 \\ \hline
$\gamma=10$ dB  & - & - & 1.25 & 1.27 & 1.32 & 1.39 \\ \hline
$\gamma=20$ dB  & - & - & 1.25 & 1.28 & 1.34 & 1.42 \\ \hline
\end{tabular}
\end{table}

\section{Conclusions}
\label{sec:Conclusion}
In this paper, coordinated multicast beamforming algorithms were proposed for a multi-cell multigroup network, where each BS sends independent sets of common data to distinct groups of users. The optimization objective is to minimize the total transmission power while guaranteeing the user-specific quality of service constraints. This non-convex problem is approximated as a convex one via the SDR method. In the case of higher rank solution, the Gaussian randomization method is used to provide feasible rank-one beamformers.
In addition to a centralized approach, we proposed a novel primal decomposition-based distributed algorithm which relies only on local CSI and limited backhaul information exchange.
The numerical results showed that the proposed beamforming coordination is beneficial compared to the conventional transmission schemes. The results also demonstrated that the distributed algorithm obtains performance close to that of the centralized approach even after few iterations.

\section*{Acknowledgements}
This work was partially supported by the National Research Fund, Luxembourg, under the projects $\mathrm{SATSENT}$ and $\mathrm{SEMIGOD}$, and by the European Commission, H2020, under the project $\mathrm{SANSA}$.


\begin{small}
\bibliographystyle{IEEEbib}
\bibliography{strings_all}
\end{small}

\end{document}